\documentclass[conference]{IEEEtran}
\ifCLASSINFOpdf
  \usepackage[pdftex]{graphicx}
\else
  \usepackage{graphicx}
\fi
\usepackage{amsmath}
\begin{document}
%
\title{An in-silico study of travel time estimation \\ using Eikonal equation with finite element method in ultrasound imaging}

\author{\IEEEauthorblockN{Feixiao Long}
\IEEEauthorblockA{Software group \\ eSonic Image (Beijing) Co., Ltd. \\
Beijing, P. R. China\\
Email: lf832003@gmail.com}
\and
\IEEEauthorblockN{Weiguang Zhang}
\IEEEauthorblockA{Department of Automation \\ Tsinghua University \\
Beijing, P. R. China\\
Email: zwg22@mails.tsinghua.edu.cn}
}


%


\maketitle

\begin{abstract}
Estimating the travel time of ultrasound in an inhomogeneous medium is crucial for high-quality imaging, as with an accurate estimate of the distribution of speed of sound, phase aberration, which is normally viewed as one of the reasons for image degradation, can be corrected by multiple methods. In order to estimate the ultrasound travel time in an inhomogeneous medium, the approach using solutions to the eikonal equation is attractive, since it can be easily integrated into a conventional delay-and-sum beamformer. In our manuscript, we first propose employing the finite element method to solve the eikonal equation in ultrasound imaging, which is easy to extend to a high-order numerical scheme and potentially runs faster on GPU. Through a series of in-silicon experiments, our results are comparable to those of fast marching methods.          
\end{abstract}


%
\IEEEpeerreviewmaketitle

\section{Introduction}

 
Phase aberration, often viewed as one of the main reasons for ultrasound image degradation, is usually caused by variations in the speed of sound (SoS) in the medium \cite{Aberration}. Generally speaking, the first critical step to correct phase aberration is to accurately estimate the distribution of the SoS in an inhomogeneous medium \cite{SoSConvex}, \cite{SoS} and the second step is to calculate the ultrasound travel time, which could be used in the beamformer, between the imaging points and the transducer elements using the SoS map acquired in the first step. In our work, we focus mainly on the second step. Although diffraction-based methods \cite{FourierSplit} can also perform phase aberration correction, considering that the delay-and-sum (DAS) beamformer is still widely used especially in commercial devices, herein the DAS beamformer is preferred, which can easily incorporate the corrected travel time. Given the distribution of SoS, one of the promising methods for calculating the corrected travel time is through the eikonal equation. Therefore, in our study, the question of how to obtain a more flexible and efficient solution to the eikonal equation is mainly investigated. 

Usually, especially in ultrasound imaging \cite{Aberration}, \cite{Aberration1}, the widely used method to solve the eikonal equation is the fast marching method (FMM) \cite{FMM}. However, as pointed out in \cite{LaplacianEikonal}, when performing domain decomposition in the 3D region with parallel computing, the key characteristic (heap data structure) of FMM cannot be maintained effectively. In addition, if the above-mentioned solvers are based on the Godunov upwind finite difference method (FDM) \cite{FIMGPU}, \cite{FIM}, the flexibility to deal with domains of arbitrary shapes is limited, since FDM is generally difficult to implement for curved boundary. Therefore, in our manuscript we propose to employ the finite element method (FEM) to solve the eikonal equation based on the following considerations. First, it is known that FEM can be easily run on a GPU or scaled to large high-performance computers \cite{Fenicsx}. Second, it is common to employ unstructured grids in FEM, such as triangular or tetrahedron-based meshes, which is helpful when dealing with domains with complex boundaries \cite{Radiative}. Third, it is more advanced and easier with FEM to construct a high-order numerical scheme, which increases the accuracy of the algorithm in general. Furthermore, compared to more specialized FMM, more general FEM software packages, such as FEniCSx \cite{Fenicsx}, can be easily acquired. 

The main contribution of our manuscript is: we develop a complete numerical scheme of continuous Galerkin (CG) FEM to solve the eikonal equation in polynomial space $Q_1$. Furthermore, to the best of our knowledge, this is the first time that FEM has been used to numerically solve the eikonal equation in \textbf{ultrasound imaging} for a linear probe. In addition, our Matlab code was preliminarily tested in the experiments.       


\section{Method}
\subsection{Theory}
The eikonal equation shown below is investigated in our method, 
\begin{equation}
\label{Eikonal}
    \begin{split}
        \| \nabla \tau(x, z) \|^{2} &= \frac{1} {c^{2}(x, z)} \\
        \tau(x, z) = 0, \quad &(x, z) \in \Gamma_{D}        
    \end{split}
\end{equation}
in which, $\tau$ represents the time of flight between point $(x, z)$ and one probe piezo-element located at $(x_{i}, z=0), i = 0, ..., N - 1$, which will be further represented as the Dirichlet boundary (see below). $N$ denotes the number of piezo-elements in the linear ultrasound probe. $c(x, z)$ denotes the distribution of SoS within the medium, which may be inhomogeneous. $\nabla$ is the gradient operator. The Dirichlet boundary condition (denoted as $\Gamma_{D}$) is set as: on a particular piezo-element $i$, the travel time is $0$. Theoretically, the eikonal equation can accurately model the propagation time of ultrasound within the refractory medium \cite{Seism}. 

To apply the FEM to solve the eikonal equation, inspired by \cite{FEMpipeline}, a viscous regularization term is added to the left-hand side of Eq. \ref{Eikonal} as shown below, 
\begin{equation}
\label{ViscosityEikonal}
    -\beta \Delta \tau(x, z) + \| \nabla \tau(x, z) \|^{2} = \frac{1}{c^{2}(x, z)}
\end{equation}
$\beta$ is a small positive number, which is related to the mesh size in general. $\Delta$ represents Laplacian operator. Intuitively, when the viscosity term approaches $0$, Eq. \ref{ViscosityEikonal} approximates to Eq. \ref{Eikonal}. Using the viscous regularized eikonal equation, the standard CG-FEM method can be applied, as briefly shown below. For simplicity, we use $f(x, z)$ to denote $\dfrac{1}{c^2(x, z)}$ for the following discussion. 

Let $\Omega = [-\frac{L}{2}, \frac{L}{2}] \times [0, D]$ ($L$ is the length of the linear probe and $D$ represents the depth of detection). We define the trial and test function space as following shown respectively, 
\begin{equation}
    \begin{split}
        V &= \{v \in H^{1}(\Omega) \, \big| \, v = 0 \; \text{on} \; \Gamma_{D} \} \\
        \hat{V} &= \{v \in H^{1}(\Omega) \, \big| \, v = 0 \; \text{on} \; \partial \Omega \} 
    \end{split}
\end{equation}
in which $H^{1}(\Omega)$ is Hilbertian Sobolev space. Multiplying both sides of Eq. \ref{ViscosityEikonal} with test function $v \in \hat{V}$, and integrating,  
\begin{equation}
\label{Integrationform}
    \big( -\beta \Delta \tau(x, z) + \| \nabla \tau(x, z) \|^{2}, v(x, z) \big) = \big( f(x, z), v(x, z) \big)
\end{equation}
Here, $(\cdot, \cdot)$ represents the inner product in $\hat{V}$, for scalar and vector-valued (see below) functions, defined as
\begin{align*}
    (u, v) &= \iint_{\Omega} uv \mathrm{d}x \mathrm{d}z \\
    (\mathbf{u}, \mathbf{v}) &= \iint_{\Omega} \mathbf{u} \cdot \mathbf{v} \mathrm{d}x \mathrm{d}z
\end{align*}
Performing integration by parts of the first term of left-hand side in Eq. \ref{Integrationform}, we can get (omitting the dependence of $(x, z)$).
\begin{equation}
\label{Integrationbyparts}
     (\beta \nabla \tau, \nabla v) + (\| \nabla \tau \|^{2}, v) = (f, v)
\end{equation}
Using Newton's method to solve above non-linear equation and for the $k$th iteration, letting $\tau = \tau_{0}^{k} + \delta \tau^{k}$, Eq. \ref{Integrationbyparts} can be reformulated as 
\begin{equation}
\label{Newtonexpression}
    \begin{split}           
        (\beta \nabla \tau_0^{k}, \nabla v) + (\beta \nabla \delta \tau^{k}, \nabla v) + (\| \nabla \tau_0^{k} + \nabla \delta \tau^{k} \|^{2}, v) = (f, v)
    \end{split}
\end{equation}
Expanding the third term of Eq. \ref{Newtonexpression} and ignoring the higher-order term of $\nabla \delta \tau^{k}$, we finally get the following form as 
\begin{equation} 
    \begin{split}
    (\beta \nabla \tau_0^{k}, \nabla v) + (\beta \nabla \delta \tau^{k}, \nabla v) + & \\ 
    (\nabla \tau_0^{k} \cdot \nabla \tau_0^{k}, v) + 2(\nabla \tau_0^{k} \cdot \nabla \delta \tau^{k}, v) &= (f, v)
    \end{split}
\end{equation}
Herein, Eq. \ref{Eikonal} is eventually transformed into the following weak form: For the $k$th iteration, find $\delta \tau^{k} \in V$, such that
\begin{equation}
\label{Weakform}
    a(\delta \tau^{k}, v) = L(v), \; \forall v \in \hat{V}
\end{equation}
in which, 
\begin{align*}
    a(\delta \tau^{k}, v) &= (\beta \nabla \delta \tau^{k}, \nabla v) + 2(\nabla \tau_0^{k} \cdot \nabla \delta \tau^{k}, v), \\
    L(v) &= (f, v) - (\beta \nabla \tau_0^{k}, \nabla v) - (\nabla \tau_0^{k} \cdot \nabla \tau_0^{k}, v)
\end{align*}

\subsection{Implementation} 
If using third-party software such as FEniCSx, Eq. \ref{Weakform} can be directly transformed into code. However, for our in-house code, several more details are necessary to translate Eq. 
\ref{Weakform}. Taking into account the nature of the linear probe, it is appropriate to use the $Q_1$ polynomial space (with the form $\phi = a_0 + a_1x + a_2z + a_3xz$) and the corresponding rectangular mesh. Here, we would like to emphasize that our method can be easily adapted to a common triangular mesh and in that case, $P_1$ polynomial space will be used. Moreover, using the $P_1$ space will be investigated in our future work since the number of nodes for the triangular mesh can be set lower than the corresponding rectangular mesh. 

In summation, the following equation is used to approximate $\delta \tau^{k}(x, z)$, namely
\begin{equation}
    \delta \tau^{k} = \sum_{i = 0}^{N_{n}} \delta \alpha_i^{k} \phi_i, \; \phi_i \in Q_1
\end{equation}
where $N_n$ represents the number of nodes in rectangular mesh. Either using numerical quadrature or directly computing the integral terms (because of a simple $Q_1$ space) in Eq. \ref{Weakform} can be easily implemented when building the stiffness and mass matrix. Although not necessary, the standard reference rectangle ($[0, 1] \times [0, 1]$) is still used in our integral computation (only some extra Jacobians are evaluated) for compatibility with the use of the $P_1$ space. We follow the regular Newton's method (set $\tau_0^{k + 1} = \tau_0^{k} + \delta \tau^{k}$) to iteratively solve Eq. \ref{Weakform} until the maximum norm of the solution vector $\delta \tau^{k}$ is less than some predefined threshold, that is, $10^{-6}$ in our experiments. Note that, in general, the coefficient matrix corresponds to the bilinear form in Eq. \ref{Weakform} is not symmetric, so the generalized minimum residual method with preconditioned coefficient matrix (GMRES in Matlab built-in functions) is used. 

\subsection{Accuracy}
For comparisons, FMM \cite{FMM} is employed as the benchmark and considering the true values of travel time in the domain are quite small, the accuracy of our proposed method is evaluated using pointwise relative error defined as
\begin{equation}
\label{Metric}
    \text{err} = \frac{|\tilde{\tau}_i - \tau_i|}{\tilde{\tau}_i}, \quad i=1, ..., N_{n}
\end{equation}
Here, $\tilde{\tau}$ denotes the result using FMM. With Eq. \ref{Metric}, some statistics of accuracy, such as maximum relative error or mean relative error throughout the imaging plane, can be easily calculated, as shown in the following sections. 

\section{In-silicon experiments}
For comparison, we used public data acquired from one linear probe (as shown in Fig. 1) with an inhomogeneous distribution of SoS in \cite{Aberration}. We also used the code \footnote{https://github.com/rehmanali1994/DistributedAberrationCorrection} published by the authors of \cite{Aberration} for a rigorous comparison. The size of the original SoS map is $676 \times 576$, with a resolution in the $x$ and $z$ directions of 0.070mm and 0.067mm, respectively. Note that the SoS map is from the first step as introduced in Sec. 1, and it is actually too large for the travel time estimation. There are 192 transducer elements in the linear probe and the piezo pitch is 0.2mm. Therefore, when constructing the numerical scheme, we set the coordinates of the $x$ grid the same as the position of each transducer element. To reduce the size of the coefficient matrix corresponding to Eq. \ref{Weakform}, different decimation factors are set along the $z$ direction and the final solution is obtained by interpolation back to the original resolution. For example, when the decimation factor is 3, the number of points along the $z$ direction is 192. The maximum and average relative errors are used as the metric in the simulation.   
\begin{figure}[!t]
    \centering
    \includegraphics[width=3.0in]{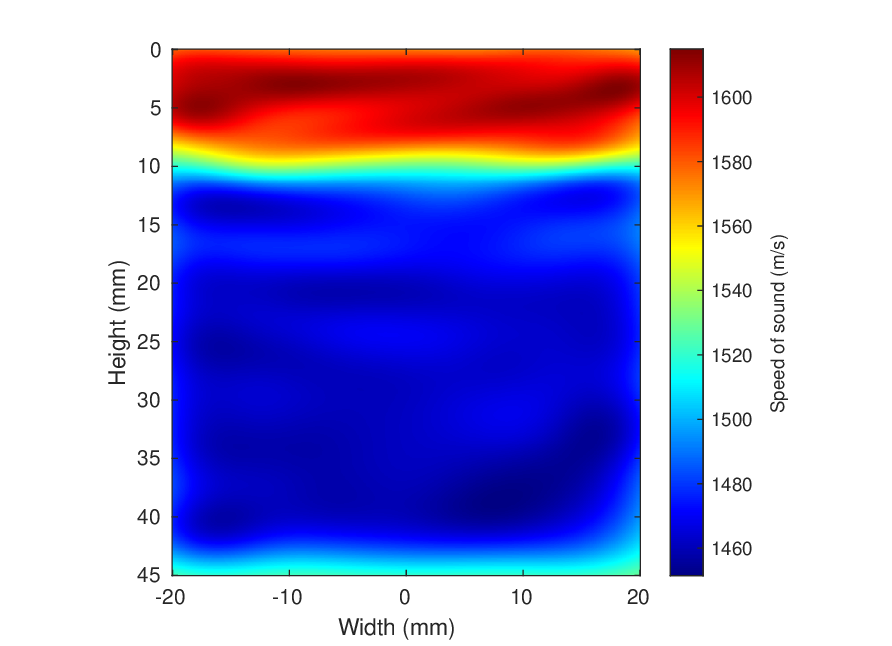}
    \caption{Reconstructed SoS map. }
    \label{Crecon}
\end{figure}

\section{Results}
For all the experiments below, we set $\beta=10^{-2}$. In the first experiment, the boundary condition is set at the transducer element with index $i=96$, that is, to estimate the flight time between each imaging point $(x, z)$ and the central element of the probe. We first show the relationship between the maximum norm of the solution vector and the number of iterations needed for solving Eq. \ref{Weakform} with different decimation factors along the $z$ direction, as shown in Fig. 2. From the figure, the convergence plot is almost the same for different decimation factors, and $\|\delta \tau^{k}\|_{\infty}$ goes below our preset threshold after about 17 iterations. This suggests that the decimation factors have little influence on the convergence of our method. However, with a larger decimation factor, the total running time is reduced.  
\begin{figure}[!t]
    \centering
    \includegraphics[width=3.0in]{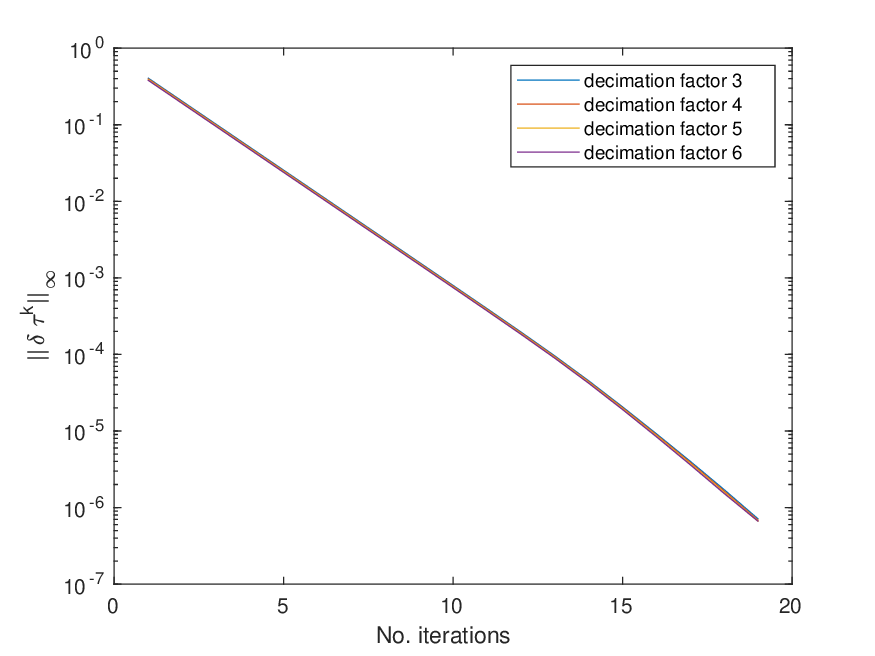}
    \caption{Convergence plot for different decimation factors along $z$ direction. }
    \label{Convergence}
\end{figure}

The mean relative error across the entire image plane for different settings is shown in Table 1. It is quite interesting to observe that, with an increasing decimation factor, the mean relative error is reduced. This is probably due to the interpolation, and we will investigate this phenomenon in our future work. The travel time is shown in Fig. 3. 
\begin{table}[!t]
\caption{Comparisons of mean relative error among different settings}
\label{Table1}
\centering
\begin{tabular}{cccc}
\hline
Decimation factor & \multicolumn{3}{c}{Boundary element position} \\
 & $i=48$ & $i=96$ & $i=144$ \\
\hline
3 & $3.6\%$ & $3.7\%$ & $3.6\%$ \\
4 & $3.3\%$ & $3.4\%$ & $3.3\%$ \\
5 & $3.0\%$ & $3.1\%$ & $3.0\%$ \\
6 & $2.8\%$ & $2.8\%$ & $2.8\%$ \\ 
\hline
\end{tabular}
\end{table}

\begin{figure}[!t]
    \centering
    \includegraphics[width=3.0in]{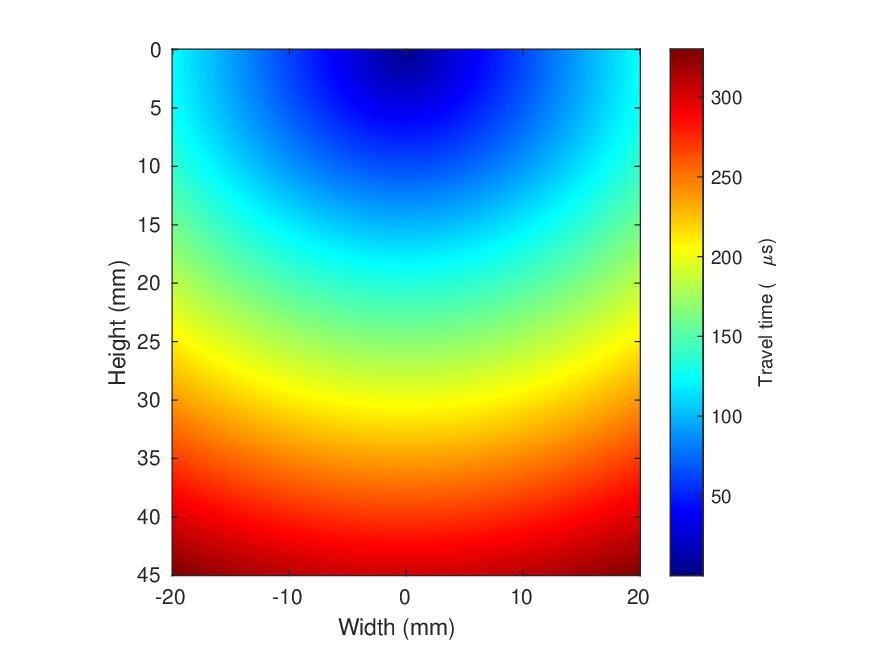}
    \caption{Travel time between the imaging points and center element in $\mu s$. }
    \label{Time}
\end{figure}

The contour between FMM and FEM based solutions (with decimation factor 6) is displayed in Fig. 4. It is clear from the figure that a large discrepancy between these two solutions only occurs at the near field of the Dirichlet boundary. This phenomenon can be anticipated due to the different boundary condition procedures of the two methods. This can also be verified from Fig. 5, which shows the relative error defined in the previous section through the whole imaging plane. From Fig. 5, for depth less than 5mm, the relative error is larger than $20\%$. With increasing depth, the relative error drops below $5\%$ quickly. The mean relative error for the entire imaging plane is only $2.8\%$. Although the maximum relative error is greater than $200\%$ near the boundary, in fact, the number of nodes with a relative error greater than $20\%$ is only 2,225, approximately $0.6\%$ of the total number of nodes in our study. 
\begin{figure}[!t]
    \centering
    \includegraphics[width=3.0in]{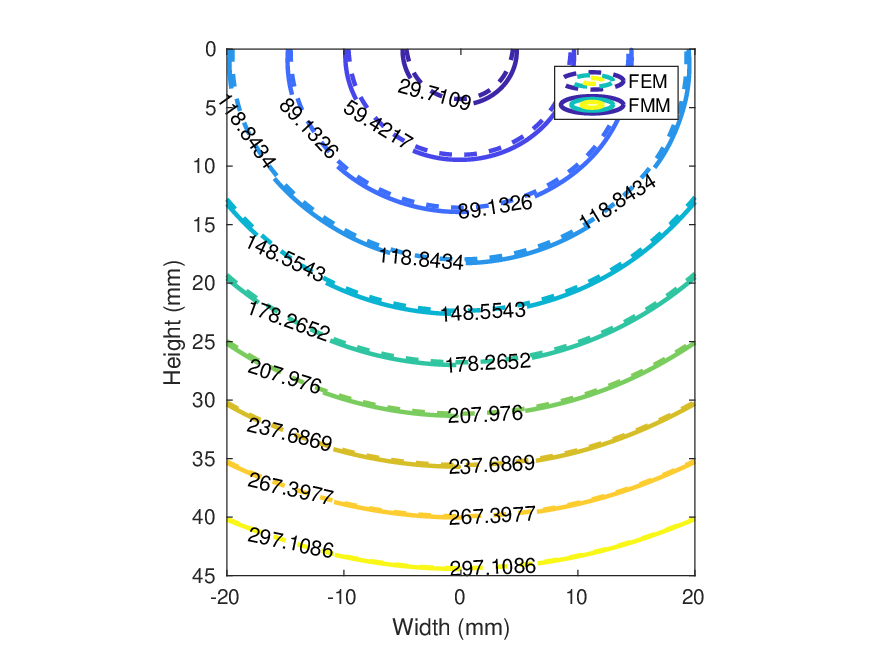}
    \caption{Contour between FMM and FEM solutions. The texts on the contour line indicate the travel time in \text{$\mu s$}. }
    \label{Contour}
\end{figure}
\begin{figure}[!t]
    \centering
    \includegraphics[width=3.0in]{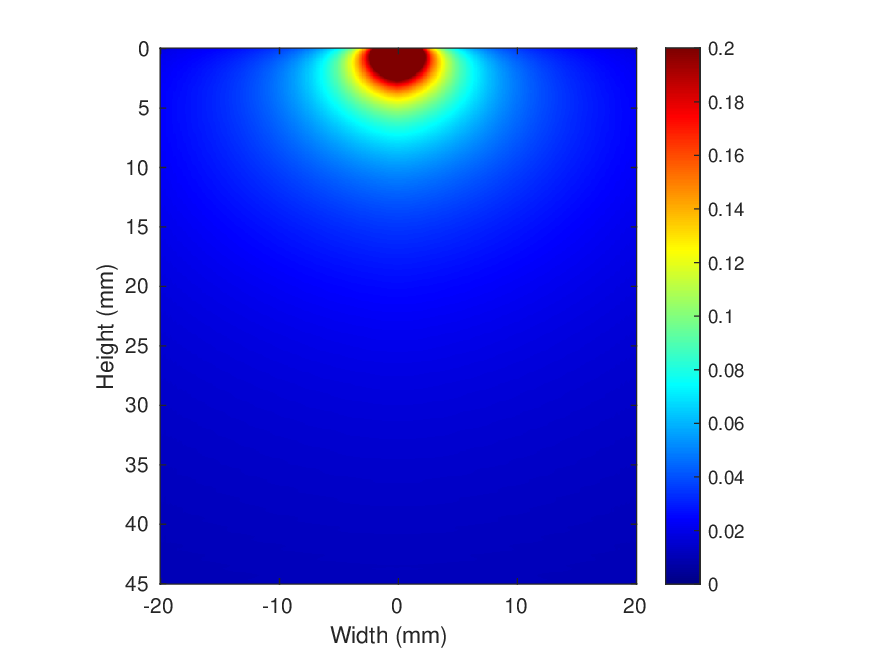}
    \caption{Relative error between FMM and FEM solutions. }
    \label{RelativeError}
\end{figure}

In Table 1, the same trends can also be observed for the boundary transducer element located on the left and right halves of the probe. Since our method has no restrictions on the position of boundary-conditioned piezo-element, it is anticipated that it will still work for these cases.

\section{Discussions}
In this manuscript, the feasibility of using FEM to solve eikonal equation with in-house Matlab code is preliminarily tested through a series of in-silicon experiments. Our solutions are comparable to the conventional FMM solutions.  

For our proposed method, it is easy to convert them to run on modern GPU. Here we can divide our proposed method into two separate parts: one is to construct the coefficient matrix, which can be done quite fast on GPU. The second is to solve the linear equations using GMRES. This part is more complex than the first part. However, with the help of software libraries such as MAGMA \cite{MAGMA} or PETSc \cite{PETSc}, it is still possible to complete the second part computation rapidly on GPU. To run the proposed methods on GPU is actually our future work.   






%

\end{document}